\documentclass[12pt,preprint]{aastex}

\shorttitle{Dual Halos in ETGs}
\shortauthors{Park \& Lee}

\begin{document}

\title{Dual Halos and Formation of Early-Type Galaxies} 

\author{ Hong Soo PARK\altaffilmark{1} \& Myung Gyoon LEE\altaffilmark{1}
\email{hspark@astro.snu.ac.kr \& mglee@astro.snu.ac.kr}
}

\altaffiltext{1}{Astronomy Program, Department of Physics and Astronomy, Seoul National University, Korea}
 
\begin{abstract}
We present a determination of the two-dimensional shape parameters of the blue and red globular cluster systems (GCSs) in a large number of elliptical galaxies and lenticular galaxies 
(early-type galaxies, called ETGs). 
We use a homogeneous data set of the globular clusters in 23 ETGs obtained from the HST/ACS Virgo Cluster Survey.
The position angles of  both blue and red GCSs show a correlation with those of the stellar light distribution, showing that the major axes of the GCSs are well aligned with those of their host galaxies.
However, the shapes of the red GCSs show a tight correlation with the stellar light distribution as well as with  the rotation property of their host galaxies, while the shapes of the blue GCSs do much less.
These provide clear geometric evidence that 
the origins of the blue and red globular clusters  are distinct and that ETGs may have dual halos: a blue (metal-poor) halo and a red (metal-rich) halo. 
These two halos show significant differences in metallicity, structure, and kinematics, indicating that they are formed in two distinguishable ways. 
The red halos might have formed via dissipational processes with rotation, while the blue halos are through accretion. 
\end{abstract} 


\section{Introduction}

Early-type galaxies (ETGs, elliptical galaxies and lenticular galaxies) appear to be morphologically simple in their images. However recent accumulating evidence shows  that they are more complex and intriguing than ever \citep{fer06,kor09,lee10a,lee10b,zhu10,cap11,gre12}. 
One of the most surprising and intriguing findings in extragalactic studies during the last two decades is a discovery that the color distribution of the globular clusters (GCs) in ETGs is bimodal, 
suggesting that there are two subpopulations: blue and  red GCs \citep{zep93,gei96,bro06,pen08}.
Blue and red GCs are generally considered to correspond to 
 metal-poor and metal-rich GCs, respectively. 

Numerous studies based on observations and simulations followed to investigate the nature and origin of these blue and red subpopulations in ETGs (\citet{bro06} and references therein). 
One of the important results is  that these subpopulations show differences in  one-dimensional spatial distribution: the radial number density profiles of the blue GCs are flatter than those of the red GCs \citep{kis97,lee98,rho04,str11,for12}. 
On the other hand, it was suggested  that the bimodal color distribution of the GCs 
can be explained by non-linear transformation between color and metallicity,
implying that it is not necessary 
to believe in the existence of two chemically distinct subpopulations in ETGs \citep{yoo06}. 
Recently several efforts were made to resolve this issue using spectroscopy and near-infrared photometry, 
but their conclusions are controversial \citep{woo10,yoo11,bla12,par12,bro12,pot13}.

Two-dimensional spatial distributions of these subpopulations can provide critical clues to settle the controversy of their origin. 
However,  two-dimensional spatial
distributions of these subpopulations in the literature show contradictory results. 
Several studies found that the ellipticity and position angle of 
both subpopulations are similar to those of the spheroids of their parent galaxies
\citep{fort01,gom04,dir05,ric12}. 
On the other hand, other studies presented evidence that  the red GCs follow the galaxy ellipticity more closely than the blue GCs \citep{kis97,lee98,for01,lee08,str11,par12a}. 
It is not yet known whether there is any common feature in the two-dimensional spatial distribution of the GCs among galaxies. 
One of the best ways to resolve this controversy 
is to use the structural shapes of the GC systems (GCSs) in a large sample of galaxies, but there has been no such studies to date.
In this paper we study the two-dimensional spatial distribution of the blue and red GCs in a large number of ETGs using a homogeneous data set 
from the Advanced Camera for Surveys Virgo Cluster Survey (ACSVCS) obtained with the Hubble Space Telescope (HST) \citep{cot04,fer06,jor09}.  

\section{Data and Method}

We used data for the position and photometry of the GCs in bright ETGs of the Virgo cluster
given in the ACSVCS \citep{jor09}. 
ACSVCS is a homogeneous $gz$ imaging survey of a  sample of 100 ETGs in Virgo with the HST \citep{cot04, fer06}. 
It  provided the most comprehensive and homogeneous photometric results on GCs in Virgo so that  it is an excellent data set for
investigating the properties of GCSs in Virgo galaxies. 
We selected 23 bright ETGs (including Es, E/S0s, and S0s)  in the ACSVCS for our analysis.
The numbers of the red GCs are mostly smaller than those of the blue GCs in each galaxy. Therefore we used only the galaxies with a large number of known red GCs ($N\ge24$). 
We use the basic properties of the sample galaxies 
$V$-band total magnitudes ($M_{V}$) from \citet{pen08}, 
effective radii ($R_{e}$), position angles (P.A.), and ellipticities ($\epsilon$) from \citet{fer06}, and
rotational parameters ($\lambda_{Re}$) from \citet{ems11}.
We adopted ellipticity and P.A. of each galaxy that are average values measured between $1\arcsec$ and effective radius in the $g$ band, although ETGs often show isophotal twisting \citep{fer06}.
$\lambda_{Re}$  is a parameter for specific angular momentum (angular momentum normalized to the mass) measured from the stars at the effective radius of the galaxy. 
It is a sensitive rotation indicator
 similar to the classical parameter, the ratio between rotational velocity and velocity dispersion ($V/\sigma_v$). 

To determine two-dimensional shape parameters of the GCSs in each galaxy,
we selected first  the bright objects with  $g_0<25.0$ mag and GC probability $P_{GC}>0.5$  in the ACSVCS GC catalog \citep{jor09}. 
Next we divided the GC sample into two groups according to their color: the blue GCs ($(g-z)_0<1.1$) and the red GCs ($(g-z)_0>1.1$).

We investigated two-dimensional shapes of the GCSs in the sample galaxies %
projected in the sky from the spatial distribution of the GCs.
The two-dimensional shapes of the GCSs projected in the images can be approximated with an ellipse.  We determined the shape parameters (ellipticity and position angle) of an ellipse for the GCSs in a galaxy,
using the method of the dispersion ellipse of the bivariate normal frequency function of position vectors 
\citep{tru53}. %
The dispersion ellipse represents a contour at which the density is 0.61 times the maximum density in the center. This method has been often used to estimate the two-dimensional shapes of galaxy clusters or GCSs in a galaxy from a catalog of sources \citep{car80,mcl94,hwa07,str11}.
We derived the errors for the parameters as 68\% 
confidence levels obtained from the bootstrapping procedure with 1000 trial.
We derived the shape parameters for three groups in each galaxy: the entire GCS, the blue GCS, and the red GCS.

\begin{figure}
\epsscale{0.8}
\plotone{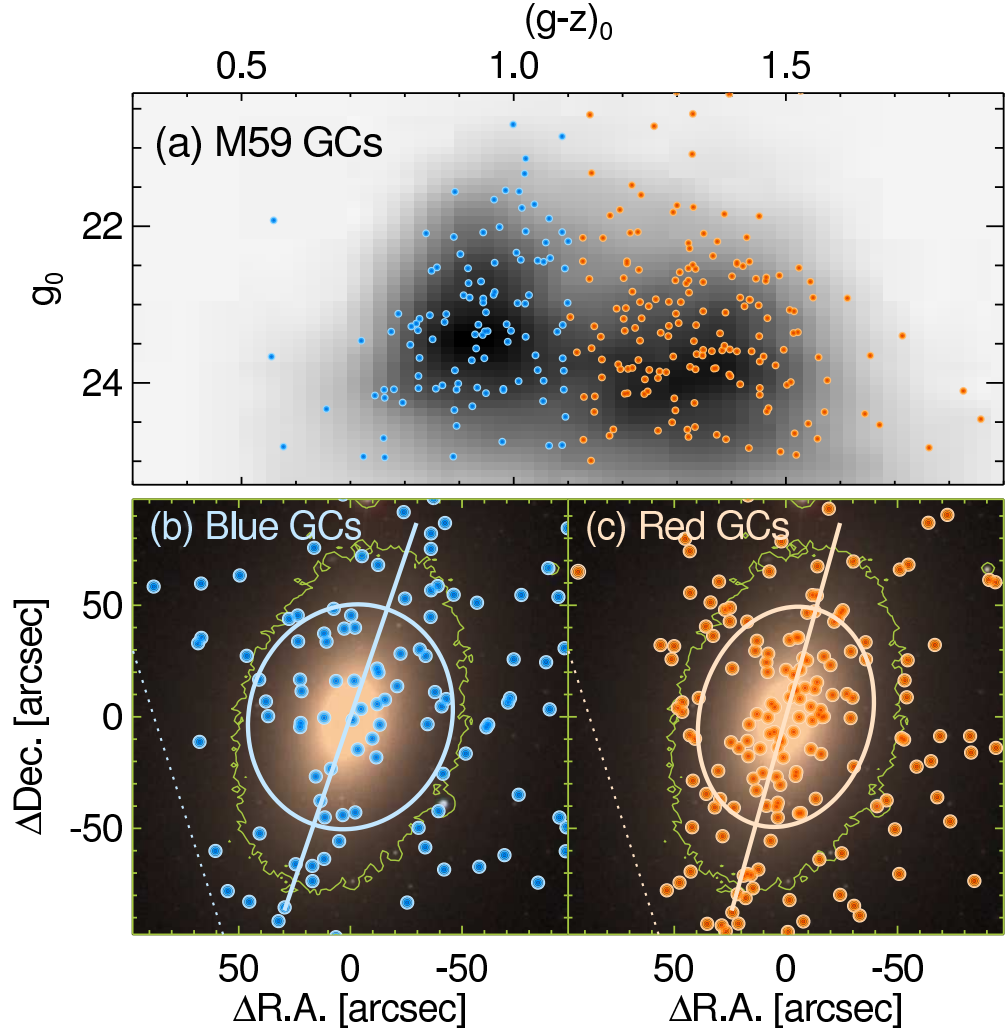}
\caption{An example for deriving the shape parameters of the GCSs in a galaxy: M59 (VCC 1903), an E5 galaxy.
(a) The color-magnitude diagram for blue and red GCs in M59.
The grey-scale image represents the number density map of the GCs in all 23 galaxies.
(b) and (c) Spatial distributions of the blue and red GCs, respectively. 
The solid ellipses and lines represent the shapes and position angles determined for the GCSs, respectively.
Green contours represent reference isophote contours overlaid on Sloan Digital Sky Survey color images  for M59.
Dotted lines represent the boundary of the HST images.
\label{fig1}} 
\end{figure}

Figure \ref{fig1} %
displays an example of our shape parameter determination  for the blue and red GCSs in an elliptical (E5) galaxy: M59. 
In Figure \ref{fig1}(a)  the color-magnitude diagram of the GCs in M59 as well as all other galaxies in the sample shows that the GCs are divided into two subpopulations: blue and red.
In Figure \ref{fig1}(b) and (c) it is seen clearly that the blue GCs are rather spread over the region, while
the red GCs show a significant central concentration elongated along 
the major axis of their host galaxy.
The ellipticity of the blue GCS we derived is $\epsilon=0.11\pm0.07$,  two times smaller than that of the red GCS, $\epsilon=0.23\pm0.07$.
The ellipticity  of the red GCS is closer to that of the stellar light distribution ($\epsilon=0.34\pm0.02$) than that of the blue GCS. The position angles of the blue and red GCS are consistent with those of their host galaxies. %
This shows that there are clear differences in the two-dimensional distribution between the blue and red GCs in this galaxy, 
and that our determination of the shape parameters is consistent with visual estimation. %

\section{Results}

\begin{figure}
\epsscale{1.0}
\plotone{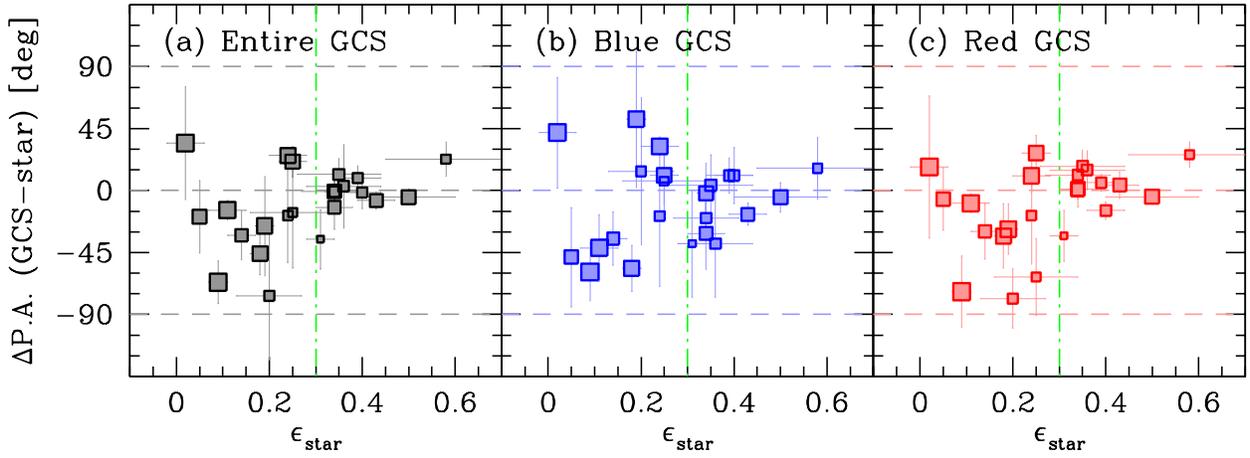}
\caption{Position angle differences ($\Delta$ P.A.) of the entire GCSs  (a), the blue GCSs  (b),  and  the red GCSs (c) with respect to their host galaxies versus ellipticities of the stellar light distribution ($\epsilon_{\rm star}$). 
The dot-dashed lines represent $\epsilon_{\rm star}=0.3$.
Symbol sizes represent the relative brightness of the host galaxies: the larger, the brighter. 
\label{fig2}}
\end{figure}

Table \ref{tab-0} lists the shape parameters (major axis length, minor axis length, position angle, and ellipticity) derived for the entire GCSs, the blue GCSs, and the red GCSs in 23 ETGs.
A comparison of the position angles of  the GCSs 
and those of the stellar light distribution in their host galaxies is displayed in Figure \ref{fig2}. 
The position angles of the entire, blue, and red GCSs show a good correlation with those of the stellar light distribution
for the elongated galaxies with  $\epsilon>0.3$, suggesting that the major axes of the GCSs are well aligned with those of their host galaxies. The large scatter in the position angle differences for less elongated galaxies
with  $\epsilon< 0.3$ is due to the difficulty in determining the position angles for small ellipticity.
This result is consistent with the results in \citet{wan13}. 
Using the same data as in this study, Wang et al. (2013) studied the azimuthal variation of the number density of the red and blue GCs, and found that both systems show alignment with the major axis of their host galaxies.

\begin{figure}
\epsscale{1.0}
\plotone{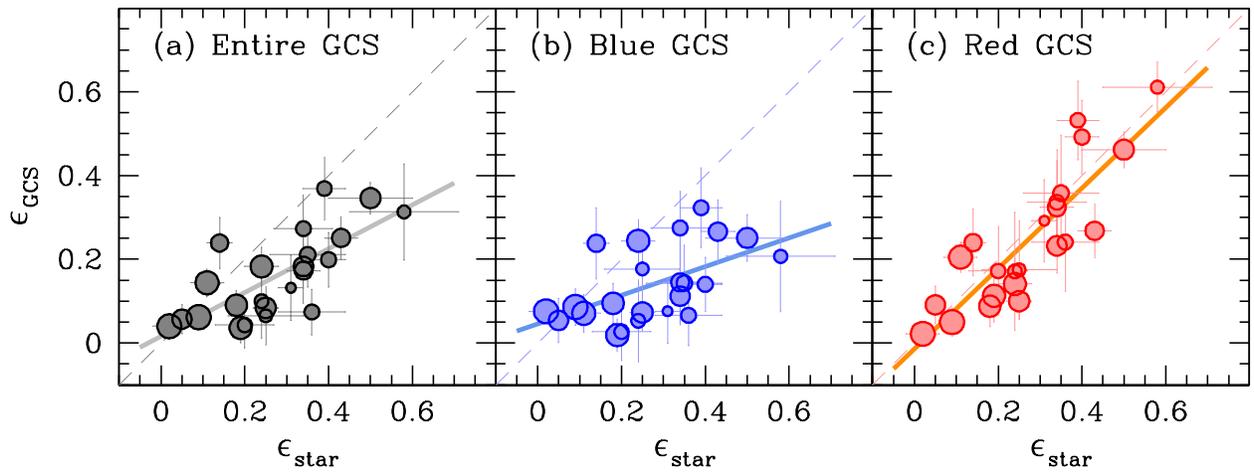}
\caption{Ellipticities of the GCSs versus those of stellar light distribution in their host galaxies. 
(a) the entire GCSs, (b) the blue GCSs, and (c) the red GCSs.
Symbol sizes are same as Figure 2.
The dashed lines represent one-to-one relations, and the solid lines represent linear fits. 
\label{fig3}} 
\end{figure}

Figure \ref{fig3} displays a comparison of the ellipticities of the GCSs we derived and those of the stellar light distribution. %
The ellipticity of the entire GCSs ($\epsilon({\rm GCS})$) in Figure \ref{fig3}(a) 
shows a weak correlation with that of the stellar light distribution  ($\epsilon({\rm star})$).
Linear fits yield $\epsilon({\rm GCS})=0.523(\pm0.108)\epsilon({\rm star})+0.016(\pm0.033)$  and rms=0.069. 
If we examine the subpopulations separately, 
the blue GCSs and the red GCSs show 
a stark contrast with respect to the stellar light distribution (Figure  \ref{fig3}(b) and (c)).
The ellipticity of the red GCSs shows a strong correlation 
with that of the stellar light distribution, 
while the ellipticity of the blue GCSs shows a much weaker correlation (Spearman correlation coefficients are 0.87 for the red GCSs and  0.56 for the blue GCSs).
We derive from linear fits,
$\epsilon({\rm BGCS})=0.342(\pm0.116)\epsilon({\rm star})+0.046(\pm0.036)$ and rms=0.074 for the blue GCSs, 
and 
$\epsilon({\rm RGCS})=0.959(\pm0.126)\epsilon({\rm star})-0.014(\pm0.039)$ and rms=0.080 for the red GCSs.
Thus  the slope for the red GCSs 
is close to one, while that for the blue GCSs 
is much smaller than one.
These results show 
that the spatial distributions of the red GCs follow closely those of the stars in their host galaxies, 
while those of the blue GCs do much less.

\begin{figure}
\epsscale{1.0}
\plotone{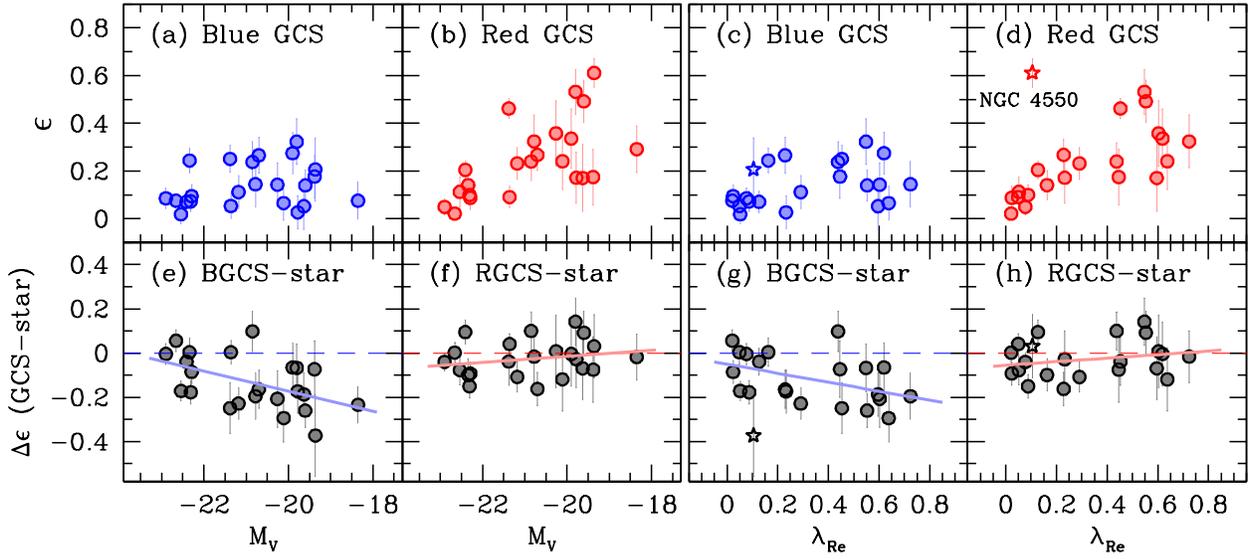}
\caption{Ellipticities ($\epsilon$) of the GCSs versus $V$-band total magnitudes ($M_V$) and rotation parameters ($\lambda_{Re}$) of their host galaxies.
(a), (b) Ellipticities of the blue and red GCSs 
versus $M_V$. 
(c), (d) Ellipticities of the blue and red GCSs 
versus  $\lambda_{Re}$. 
(e), (f) Ellipticity differences between the GCSs and stellar light distributions versus  $M_V$ for the blue and red GCSs, respectively. 
Solid lines represent linear fits. 
(g), (h) Ellipticity differences between the GCSs and stellar light distributions versus  $\lambda_{Re}$ 
for the blue and red GCSs, respectively. 
One outlier with a large ellipticity and small $\lambda_{Re}$ (open starlet) is  NGC 4550 (VCC 1619). It is a highly elongated (E7/S0) galaxy, but is known to have counter-rotating components resulting in low rotation   \citep{ems11}.
\label{fig4}} 
\end{figure}

Figure \ref{fig4}(a), (b), (e), and (f) display the ellipticity of 
the blue GCSs, the red GCSs, and the stellar light distribution, 
as a function of  the $V$-band total magnitude ($M_V$) of their host galaxies. 
The ellipticity of the blue GCSs changes little 
depending on the total magnitude. 
In contrast, the ellipticity of the red GCSs increases, on average, as their host galaxies get fainter. 
In Figure \ref{fig4}(e) and (f)
  linear fits for the ellipticity differences yield
$\Delta\epsilon({\rm BGCS-star})=-0.045(\pm0.017)M_V-1.079(\pm0.365)$ with rms=0.102, and \\
$\Delta\epsilon({\rm RGCS-star})=0.014(\pm0.013)M_V+0.260(\pm0.280)$ with rms=0.079.
The ellipticity differences between the red GCS and stellar light distribution are fit by a linear relation with an almost zero slope, while those between the blue GCS and stellar light distribution are fit by a linear relation with a large slope. %
This shows that the ellipticities of the red GCSs follow tightly those of the stellar light distribution, irrespective
of the luminosity of their host galaxies.

We investigate the relation between the ellipticity of the GCSs and the rotational parameter of their host galaxies ($\lambda_{Re}$) \citep{cap11,ems11} in Figure \ref{fig4}(c), (d), (g), and (h).
Figure \ref{fig4}(c) and (d) show that  the ellipticity of the red GCSs  
shows a strong correlation with $\lambda_{Re}$, 
while that of the blue GCSs does not
(Spearman correlation coefficient for the red GCSs is 0.80, 
two times larger than that for the blue GCSs, 0.37).
Linear fits for the ellipticity differences in Figure \ref{fig4}(g) and (h) yield 
$\Delta\epsilon({\rm BGCS-star})=-0.204(\pm0.094)\lambda_{Re}-0.050(\pm0.038)$ with rms=0.97, and 
$\Delta\epsilon({\rm RGCS-star})=0.080(\pm0.079)\lambda_{Re}-0.055(\pm0.032)$ with rms=0.081.  
The ellipticity differences between the red GCS and stellar light distribution are fit by a linear relation with an almost zero slope, while those between the blue GCS and stellar light distribution are fit by a linear relation with a large slope. 
This result suggests that the red GCSs  may follow closely the kinematics of the stars,  irrespective of the rotation parameters of their host galaxies, while the blue GCSs do not.
This is consistent with the result
that the velocity dispersion and rotation of the red GCSs shows a stronger
correlation with that of the stellar light than that of the blue GCSs
in several ETGs \citep{lee10b, pot13}.

We checked the variation of the ellipticity differences between GCSs and stellar light distribution depending on the color range of GCs
  using only 13 galaxies with a large number of GCs ($N\ge24$ in each color bin).
We calculated the mean values of the ellipticity differences  in the moving color bins with width of $(g-z)_0=0.3$ and step of $(g-z)_0=0.05$.
The mean values of the ellipticity differences are constant at 
$\Delta\epsilon\sim-0.1$ for $(g-z)_0=0.8$ to 1.1, 
increase to $\Delta\epsilon\sim0.0$ at $(g-z)_0\approx1.4$, 
and become constant again thereafter.
Thus the ellipticity differences  between GCSs and stellar light distribution show a hint for discontinuity, indicating that the blue and red GCSs are two distinct components.

\section{Discussion}

\subsection{Dual Halos in ETGs}

The differences in the shapes of the blue and red GCSs found in this study show  that the origin of the red GCs in ETGs is similar to that of the stars, but different from that of the blue GCs. This indicates that they may represent two separate components in the structure of each galaxy. It is also supported by the fact that the radial number density profiles of the GCs in ETGs show two separate components (e.g., \citet{lee98,for12}). 
From these we conclude that there may be dual halos in these galaxies: a blue (metal-poor) halo and a red (metal-rich) halo, 
in contrast to the traditional view that typical ETGs consist of a single spheroidal component (a single halo).
These two halos have significant differences in several aspects.
The red halos include red GCs as well as red (metal-rich) stars contributing significantly to the luminosity of their host galaxies, 
while the blue halos include blue GCs and blue (metal-poor) stars that are barely visible in typical optical images of ETGs because of their low number density \citep{har07}. 
Typical optical and infrared images of ETGs show only the red halos.
The blue halos are metal-poor, while the red halos are metal-rich.
The red halos are spatially more elongated and centrally concentrated than the blue halos.
 The blue halos are much more extended than 
the red halos, as seen  in the GC map of the Virgo cluster \citep{lee10a}. 
The red halos may be rotating, while the blue halos do little \citep{lee10b, pot13}.
It is noted that the dual nature of the GCSs in ETGs is similar to that of the dual stellar halos in the Milky Way Galaxy 
\citep{car07,bee12,mcc12}. 

The similarity in the shapes of the red GCSs and the stellar light distribution  implies also that the red GCSs and stars may
share their history 
from their birth in ETGs. In addition, the radial color profiles of the red GCs in ETGs are known to be similar  to those of the stellar light \citep{lee98,bro06,ric12,par12a}. 
These results suggest that the red GCs and stars in ETGs are formed in the same place and follow similar dynamical evolution.
The large range of their ellipticity indicates that 
the ETGs have a  diverse  rotational property,  consistent with the
results from stellar kinematics \citep{cap11,ems11}.
These also suggest that the red halos might have formed mostly via various gaseous  mergers \citep{kho11} and/or dissipative collapse of the rotating proto-disks \citep{mcc12}. 
The fact that the red GCSs have on average higher values of ellipticity than the blue GCSs indicates that the red halos might have been involved with major gaseous mergers in the later epochs than the blue halos.

On the other hand, the small ellipticity and blue color of the blue GCSs and their independence from stellar light distribution show that they were formed separately from stars in the main body of ETGs. 
The blue GCs might have  formed mostly in low-mass dwarf galaxies and they were accreted later to their current host galaxies via mass assembly during the growing phase of ETGs \citep{cot98,cot00,bro06,lee10b,par12,ton13}. 
Independence of the ellipticity of the blue GCSs from the luminosity of their host galaxies indicates that their formation mechanism may not depend much on the galaxy mass.

The formation history of these two halos is consistent with the two-phase models for galaxy formation based on numerical simulations \citep{ose10}. 
Previous simulations on ETGs are focused on explaining mostly the stellar light distribution in ETGs. %
However, future simulations need to include both the blue and red GCSs as well as  the stellar light distribution for better understanding how ETGs formed. 
Our results on the difference in the two-dimensional shapes of the blue GCSs and the red GCSs in ETGs,
as well as kinematics of GCSs in ETGs \citep{lee10b,pot13} and bimodal metallicity distributions of GCs in some ETGs \citep{woo10,par12,bro12}, 
are against the scenario suggested by \citet{yoo06,yoo11} that ETGs do not have to possess two distinct sub-populations of GCs.

\subsection{Shapes of the Stellar Halos in ETGs}

We checked the ratio of the ellipticity of the blue and the red GCSs to the ellipticity of the stellar light distribution as a function of major axis length 
of the GCSs ($a({\rm GCS})$) divided by 
effective radii of their host galaxies ($R_{\rm e}$). 
The values of $a({\rm GCS})/R_{\rm e}$  
indicate the relative sizes of the measured GCSs with respect to the sizes of the stellar light distribution of their host galaxies.
The ellipticity ratios of the blue and red GCSs vary little depending on $a({\rm GCS})/R_{\rm e}$.
The mean value of the ratio for the blue GCS is $\epsilon({\rm BGCS})/\epsilon({\rm star})=0.51\pm0.30$, which is much smaller than that for the red GCS ($0.95\pm0.41$).
This result shows that the two-dimensional shapes of the blue GCSs are rounder than those of the red GCSs regardless of galactocentric radius.
If the kinematics of the blue GCSs in the outer region of their host galaxy are combined with the shapes of the GCSs, it will provide a strong constraint to study the distribution and shapes of the dark matter halos \citep{sac99,mo10}.  

Previous studies found evidence for the existence of metal-poor halo stars as well as metal-rich halo stars in some ETGs.
\citet{har07} found, from the deep photometry of the halo stars in a remote field located at 33 kpc from the center of NGC 3379 (E1), that the metallicity distribution of the stars is extremely broad and flat, requiring a need for a distinct two-stage chemical evolution model. From this they also predicted that most large ETGs will host diffuse, very low-metallicity halo components. However, their field covered only a tiny fraction of the halo so that their results could not tell about the 
geometric shape of the halo. Our results suggest that the metal-poor halo in this galaxy is less elongated than the red halo.


\acknowledgments 
The authors are grateful to Prof. Paul Hodge, and Drs. Ho Seong Hwang and Narae Hwang for 
reading carefully the original manuscript and  their useful comments.
This work was supported by the National Research Foundation of Korea (NRF) grant
funded by the Korea Government (MEST) (No.2013R1A2A2A05005120). %



\begin{deluxetable}{rrrrrr}
\tablewidth{0pc} 
\tablecaption{Shape parameters for the GCSs in the 23 galaxies\label{tab-0}}
\tablehead{ 
\colhead{VCC} & \colhead{$N$(GC)} & \colhead{$a$} & \colhead{$b$} &
\colhead{P.A.} & \colhead{$\epsilon$} \\
\colhead{} & \colhead{} & \colhead{(arcsec)} & \colhead{(arcsec)} &
\colhead{(deg)} & \colhead{} 
}
\startdata 
\multicolumn{6}{c}{\underbar{Entire GCSs}}\\
 1226 &  635 & $   54.2\pm  1.1 $ & $   50.8\pm  1.0 $ & $   93.6\pm15.3 $ & $  0.062\pm0.028 $ \\ 
 1316 & 1468 & $   52.1\pm  0.7 $ & $   50.0\pm  0.7 $ & $   80.5\pm17.8 $ & $  0.040\pm0.019 $ \\ 
 1978 &  670 & $   52.5\pm  1.3 $ & $   45.0\pm  1.1 $ & $   82.2\pm05.7 $ & $  0.144\pm0.033 $ \\
 1903 &  280 & $   50.6\pm  2.0 $ & $   41.3\pm  1.8 $ & $  162.7\pm07.1 $ & $  0.183\pm0.052 $ \\
 .... & & & & & \\
\multicolumn{6}{c}{\underbar{Blue GCSs}}\\ 
 1226 &  232 & $   55.3\pm  1.8 $ & $   50.6\pm  1.8 $ & $  101.1\pm21.1 $ & $  0.086\pm0.043 $ \\ 
 1316 &  530 & $   55.0\pm  1.1 $ & $   50.9\pm  1.1 $ & $   88.1\pm15.2 $ & $  0.076\pm0.027 $ \\ 
 1978 &  234 & $   52.1\pm  1.7 $ & $   48.5\pm  1.5 $ & $   54.7\pm23.7 $ & $  0.071\pm0.045 $ \\
 1903 &  108 & $   51.0\pm  2.6 $ & $   45.2\pm  2.5 $ & $  161.2\pm22.0 $ & $  0.112\pm0.069 $ \\ 
 .... & & & & & \\
\multicolumn{6}{c}{\underbar{Red GCSs}}\\   
 1226 &  403 & $   53.6\pm  1.3 $ & $   50.9\pm  1.2 $ & $   86.7\pm26.0 $ & $  0.050\pm0.034 $ \\ 
 1316 &  938 & $   50.4\pm  0.8 $ & $   49.3\pm  0.9 $ & $   63.2\pm35.8 $ & $  0.022\pm0.021 $ \\ 
 1978 &  436 & $   53.2\pm  1.5 $ & $   42.3\pm  1.3 $ & $   87.1\pm05.0 $ & $  0.205\pm0.036 $ \\
 1903 &  172 & $   50.1\pm  2.6 $ & $   38.5\pm  2.4 $ & $  164.4\pm07.2 $ & $  0.232\pm0.065 $ \\
 .... & & & & &  
\enddata
\tablecomments{This table is available in the online journal. 
}
\end{deluxetable}
\clearpage




\clearpage

\clearpage

\clearpage

\clearpage

\end{document}